\documentclass[aps,prl,twocolumn,lengthcheck]{revtex4}
\usepackage{amsfonts, amssymb, amsmath,graphicx}
\usepackage{subfigure}
\usepackage{comment}
\def\simge{\mathrel{%
         \rlap{\raise 0.511ex \hbox{$>$}}{\lower 0.511ex \hbox{$\sim$}}}}
\def\simle{\mathrel{
         \rlap{\raise 0.511ex \hbox{$<$}}{\lower 0.511ex \hbox{$\sim$}}}}

\begin{document}

\title{Condensation transition of ultracold Bose gases with Rashba spin-orbit coupling}

\author{Tomoki Ozawa}
\author{Gordon Baym}
\affiliation{
Department of Physics, University of Illinois at Urbana-Champaign, 1110 W. Green St., Urbana, Illinois 61801, USA
}%

\date{\today}

\def\del{\partial}
\def\p{\prime}
\def\simge{\mathrel{%
         \rlap{\raise 0.511ex \hbox{$>$}}{\lower 0.511ex \hbox{$\sim$}}}}
\def\simle{\mathrel{
         \rlap{\raise 0.511ex \hbox{$<$}}{\lower 0.511ex \hbox{$\sim$}}}}
\newcommand{\feynslash}[1]{{#1\kern-.5em /}}

\begin{abstract}
We study the Bose-Einstein condensate phase transition of three-dimensional ultracold bosons with isotropic Rashba spin-orbit coupling.
Investigating the structure of Ginzburg-Landau free energy as a function of the condensate density,
we show, within the Bogoliubov approximation, that the condensate phase transition is first order with a jump in the condensate density.
We calculate the transition temperature and the jump in the condensate density at the transition for large spin-orbit coupling, where the transition temperature depends linearly on the density of particles.
Finally, we discuss the feasibility of producing the phase transition experimentally.
\end{abstract}

\maketitle

The recent experimental realization of artificial non-Abelian gauge fields in ultracold atomic gases~\cite{Lin2011,Williams2012,Wang2012arXiv} has opened the prospect of exploring a wide range of physics that is difficult to realize or investigate in other systems.
Of particular interest is simulated Rashba-Dresselhaus spin-orbit coupling~\cite{Rashba1960,Dresselhaus1955}, for which the modified single-particle spectrum leads to novel phases and phenomena.  A notable feature of such systems is that the normal phase is not kinematically forbidden at any nonzero temperature, unlike in the usual three-dimensional Bose gases where there is no normal state below the Bose-Einstein condensation (BEC) transition temperature.  On the other hand, interparticle interactions stabilize Bose condensates at low temperatures ~\cite{Ozawa2012b,Barnett2012}.  The issue we investigate in this paper is the  phase transition at a nonzero temperature between the normal and  condensed states.

 The ground state of ultracold bosons with Rashba-Dresselhaus spin-orbit coupling is predicted, depending on the details of the interactions,  to have 
two characteristic phases: a plane-wave state, which is a BEC in a single momentum state, or a striped state, which is a BEC of two opposite momenta~\cite{Ho2011,Wang2010,Zhai2012,Ozawa2012a,Li2012}.
More exotic phases are predicted at nonzero temperatures~\cite{Jian2011} and in harmonic traps~\cite{Wu2011,Sinha2011,Hu2012,Xu2012,Ozawa2012c}.  Atoms with isotropic in-plane Rashba spin-orbit coupling have circularly degenerate single-particle energy minima; as a consequence, non-interacting bosons with such couplings do not Bose condense in three dimensions, owing to the density of states being two dimensional at low energy~\cite{Stanescu2008}.
On the other hand, as mentioned, interparticle interactions allow Bose condensation at nonzero temperatures.
Thus, the phase of the system, whether condensate or normal, is determined from energetics. 

 We focus here on spatially homogeneous ultracold bosons in three dimensions with isotropic in-plane Rashba spin-orbit coupling.  Assuming  a plane-wave condensate at low temperatures, we explore the condensate phase transition by obtaining the Ginzburg-Landau free energy as a function of the condensate density around the transition, within the Bogoliubov approximation including Hartree-Fock energies (henceforth referred to as BHF).
Calculating how the coefficients of the Ginzburg-Landau free energy vary with temperature, we conclude that the transition  between the condensed and normal phases at this level of approximation is first order.  For relatively large spin-orbit coupling, we estimate the transition temperature, which depends linearly on the density of the particles $n$.  With an increase in the spin-orbit coupling, scattering length, or particle density, it can become feasible to see the transition in realistic experimental setups.

The first order transition we find is distinct from the spurious first order phase transition one finds within BHF in usual Bose gases \cite{Holzmann2003}.  The spurious transition is driven by order parameter fluctuations which
lead to a density of particles excited out of the condensate near the transition $\sim n - \alpha n_0^{1/2}$,  which is non-analytic in the condensate fraction, $n_0$.  The first order transition is removed when correctly determining the critical behavior at the phase transition.  The relevant momentum scale there is $\sqrt{a n_0}$, where $a$ is the interparticle s-wave scattering length.  However, Rashba spin-orbit coupling introduces a second scale $\kappa$, the strength of the spin-orbit coupling (see Eq.~(\ref{H}) below); we see as a consequence that the density of excited particles is analytic in $n_0$ for $\kappa^2 \gg an_0$.   

The deep issue of whether the present transition remains first order at a higher level of approximation is beyond the scope of this paper.
On the one hand, this system is similar to 
other bosonic systems with continuously degenerate single-particle minima, such as a weak-crystallization model~\cite{Brazovskii1975} and magnon systems~\cite{Jackeli2004}, in which condensation transitions are predicted to be first order.
On the other hand, the single-particle density of states in the plane-wave phase closely resembles that in a two-dimensional Berezhinskii-Kosterlitz-Thouless system.
In finite geometry, the condensate fraction is discontinuous at the Berezhinskii-Kosterlitz-Thouless transition, but in a macroscopic system correlation corrections change the transition from first order to continuous,  as one sees from scaling arguments  \cite{bkt-holzmann}.
Independent of the precise order of the transition, our analysis provides a good approximation for the thermodynamic functions over a wide range of temperatures, except possibly in the immediate neighborhood of the transition due to the limitations of the Ginzburg-Landau formalism and mean-field theory. The thermodynamic functions we obtain are useful both theoretically and experimentally in calculating properties of the gas, e.g., dynamics~\cite{Shenoy1997}.

{\em Hamiltonian.}
We consider a system of bosons in two hyperfine (pseudospin) states labeled $a$ and $b$, with an isotropic in-plane Rashba-Dresselhaus spin-orbit coupling and an isotropic s-wave interaction, described by the Hamiltonian
\begin{align}
	\mathcal{H}
	&=
	\sum_{\mathbf{p}}
	\begin{pmatrix}
	a_\mathbf{p}^\dagger & b_\mathbf{p}^\dagger
	\end{pmatrix}
	\left[
	\frac{p^2 + \kappa^2}{2m}I + \frac{\kappa}{m} (\sigma_x p_x + \sigma_y p_y)
	\right]
	\begin{pmatrix}
	a_\mathbf{p} \\
	b_\mathbf{p}
	\end{pmatrix}
	\notag \\
	&+
	\frac{g}{2V}
	\sum_{\mathbf{p}_1 + \mathbf{p}_2 = \mathbf{p}_3 + \mathbf{p}_4}
	\left(
	a_{\mathbf{p}_4}^\dagger a_{\mathbf{p}_3}^\dagger a_{\mathbf{p}_2} a_{\mathbf{p}_1}
	\right.
	\notag \\
	&\hspace{1cm}\left.
	+
	b_{\mathbf{p}_4}^\dagger b_{\mathbf{p}_3}^\dagger b_{\mathbf{p}_2} b_{\mathbf{p}_1}
	+
	2 a_{\mathbf{p}_4}^\dagger b_{\mathbf{p}_3}^\dagger b_{\mathbf{p}_2} a_{\mathbf{p}_1}
	\right).
	\label{H}
\end{align}
As before~\cite{Ozawa2012b}, $m$ is the atomic mass; $V$ is the volume of the system; $\kappa$ is the spin-orbit coupling strength, taken to be positive; and $a_\mathbf{p}$ and $b_\mathbf{p}$ are annihilation operators of particles of momentum $\mathbf{p}$ in the pseudospin states $a$ and $b$.
The $\sigma_x$ and $\sigma_y$ are Pauli matrices between the internal states, and 
$I$ is the two-by-two identity matrix.
We assume an isotropic (constant) mean-field coupling $g$;
extending the present theory to include the effects of renormalization of the interaction~\cite{Gopalakrishnan2011,Ozawa2011,Ozawa2012a} is left as a future problem.
Diagonalization of the single-particle terms in the Hamiltonian gives two single-particle dispersion branches
$\epsilon_\pm (\mathbf{p}) \equiv \{(p_\perp \pm \kappa)^2 + p_z^2\}/2m$, 
where $p_\perp \equiv \sqrt{p_x^2+p_y^2}$,
with circularly degenerate ground states along $(p_\perp,p_z) = (\kappa, 0)$.
A previous study \cite{Barnett2012} shows that the plane-wave state, in which the condensate is made of particles with a single momentum, is the preferred ground state within the Bogoliubov approximation.
We start from the plane-wave ground state with momentum $\boldsymbol\kappa \equiv (\kappa, 0, 0)$ and consider how the transition to the normal state takes place at finite temperatures.

{\em Normal state.}
We consider the free energies of the normal and condensed states as functions of temperature $T$ and chemical potential $\mu$; at the phase transition, the chemical potentials and pressures of the normal and condensate phases must be equal.  In the normal state, the quasiparticle dispersion relation within Hartree-Fock is~\cite{Ozawa2012b}
\begin{align}
	\xi_\pm (\mathbf{p}) = \frac{(p_\perp \pm \kappa)^2 + p_z^2}{2m} - \mu + \frac32gn(\mu),
\end{align}
where the total number of particles in the normal phase $n(\mu)$, which is a function of the chemical potential $\mu$ and the temperature $T$, is self-consistently determined from the number equation
\begin{align}
	n(\mu) = \frac{1}{V}\sum_\mathbf{p} \left\{ f (\xi_- (\mathbf{p})) + f (\xi_+ (\mathbf{p})) \right\},
	\label{neqnorm}
\end{align}
where $f (x) \equiv 1/(e^{x/T} - 1)$.
For a given total density $n$, the chemical potential approaches $3gn/2$ as $T\to 0$, and,
for nonzero temperatures, $\mu < 3gn/2$.
The shift in chemical potential $\Delta \mu \equiv \mu -3gn(\mu)/2 < 0$ from its $T=0$ value
can be written in terms of the normal state density $n (\mu)$ using Eq.~(\ref{neqnorm}).  For $|\Delta \mu| \ll \epsilon_\kappa$ and $T/\epsilon_\kappa \simle 1$, the main contribution comes from $p_\perp \sim \kappa$ and $p_z \sim 0$ in $\sum f(\xi_- (\mathbf{p}))$, and we obtain 
\begin{align}
	n(\mu)
	\approx
	-\frac{m\kappa T}{2\pi}
	\ln \left( -\Delta \mu / T\right), \label{neqapprox}
\end{align}
which is essentially the mean-field result for a two-dimensional system (see Eq.~(13) of \cite{bkt-holzmann}). 
Since at fixed $\mu$, $n(\mu)$ increases with $T$, $|\Delta \mu|$ also increases with $T$. 

{\em Ginzburg-Landau free energy.}
We now determine, within BHF,  the Ginzburg-Landau free energy as a function of the condensate density $n_0$ around the condensate transition. 
Since the operator $(a^\dagger_{\boldsymbol\kappa} - b^\dagger_{\boldsymbol\kappa})/\sqrt{2}$ creates a particle in the plane-wave condensate, it is easier, as before~\cite{Ozawa2012b}, to work in the following $(-,+)$ basis:
\begin{align}
	\begin{pmatrix}
	\psi_{-,\mathbf{p}} \\ \psi_{+,\mathbf{p}}
	\end{pmatrix}
	\equiv
	\frac{1}{\sqrt{2}}
	\begin{pmatrix}
	1 & -1 \\ 1 & 1
	\end{pmatrix}
	\begin{pmatrix}
	a_\mathbf{p} \\ b_\mathbf{p}
	\end{pmatrix}.
\end{align}
The state created by $\psi^\dagger_{-,\boldsymbol\kappa}$ is macroscopically occupied.

The effective Hamiltonian in the $(-,+)$ basis within BHF is
\begin{widetext}
\begin{align}
	&\frac{1}{V}\left(\mathcal{H} - \mu \mathcal{N}\right)
	=
	-\mu n_0 + \frac{gn_0^2}{2} - g(n_-^2 + n_+^2 + n_- n_+)
	+ \frac{g n_0}{2}\frac{1}{V}\sum_{\mathbf{p} \neq \mathbf{\boldsymbol\kappa}}
	\left(
	\psi_{-, \mathbf{p}}^\dagger \psi_{-, 2\boldsymbol\kappa - \mathbf{p}}^\dagger
	+
	\psi_{-, \mathbf{p}} \psi_{-, 2\boldsymbol\kappa - \mathbf{p}}
	\right)
	\notag \\
	&+
	\frac{1}{V}\sum_{\mathbf{p}\neq \boldsymbol\kappa}
	\begin{pmatrix}
	\psi_{-, \mathbf{p}}^\dagger & \psi_{+, \mathbf{p}}^\dagger
	\end{pmatrix}
	\begin{pmatrix}
	\frac{(\mathbf{p}-\boldsymbol\kappa)^2}{2m} - \mu + g(2n_0 + 2n_- + n_+) & -i\frac{\kappa}{m} p_y
	\\
	i\frac{\kappa}{m} p_y & \frac{(\mathbf{p} + \boldsymbol\kappa)^2}{2m} - \mu + g(n_0 + n_- + 2 n_+)
	\end{pmatrix}
	\begin{pmatrix}
	\psi_{-, \mathbf{p}}
	\\
	\psi_{+, \mathbf{p}}
	\end{pmatrix},
\end{align}
\end{widetext}
where $n_0$ is the density of condensate particles and $n_-$ and $n_+$ are the densities of particles in $(-)$ and $(+)$ states that are not in the condensate.
The derivative of the free energy ${\mathcal F}(\mu, T, n_0)$ with respect to $n_0$ is
\begin{align}
	\frac{\partial \mathcal{F}}{\partial n_0}
	&=
	\frac{1}{V}\left \langle \frac{\partial (\mathcal{H} - \mu \mathcal{N})}{\partial n_0} \right \rangle
	=
	-\mu + g(n_0 + 2n_- + n_+)
	\notag \\
	&+
	\frac{g}{2V}\sum_{\mathbf{p} \neq \boldsymbol\kappa}
	\left\langle
	\psi_{-, \mathbf{p}}^\dagger \psi_{-, 2\boldsymbol\kappa - \mathbf{p}}^\dagger + 
	\psi_{-, \mathbf{p}} \psi_{-, 2\boldsymbol\kappa - \mathbf{p}}
	\right\rangle. \label{dfdn0}
\end{align}
As $n_0 \to 0$, one recovers the free energy of the normal phase  $n_-, n_+ \to n(\mu)/2$, and the last term approaches zero in this limit.
In the following, we expand the right side of (\ref{dfdn0}) as a function of $n_0$ to obtain the difference of the free energies  $\mathcal{F}$ in the condensed and normal phases.  The expansion is facilitated by using  the single-particle matrix Green's functions
with anomalous components, $\mathbf{G}(\mathbf{q},t_1-t_2)\equiv -i \langle T\left( \Psi_\mathbf{q}(t_1) \Psi^\dagger_\mathbf{q}(t_2)\right) \rangle$,
where the four-component spinor $\Psi_\mathbf{q}(t)$ is
\begin{align}
	\Psi_\mathbf{q}(t)
	&\equiv \notag \\
	&\left(
	\psi_{-,\boldsymbol\kappa + \mathbf{q}}(t), \psi^\dagger_{-,\boldsymbol\kappa-\mathbf{q}}(t),
	\psi_{+,\boldsymbol\kappa+\mathbf{q}}(t), \psi^\dagger_{+,\boldsymbol\kappa-\mathbf{q}}(t)
	\right).
\end{align}
In terms of $\mathbf{G}$, Eq.~(\ref{dfdn0}) becomes
\begin{align}
	&\frac{\partial \mathcal{F}}{\partial n_0}
	=
	-\mu + gn_0
	- gT\sum_\nu \int \frac{d^3 q}{(2\pi)^3}
	\left(
	2G_{11}(\mathbf{p},z_\nu)
	\phantom{\frac{1}{2}}
	\right.
	\notag \\
	&\left.
	+ G_{33}(\mathbf{p},z_\nu)
	+ \frac12\left(G_{21}(\mathbf{p}, z_\nu) + G_{12}(\mathbf{p}, z_\nu)\right)
	\right), \label{dfdn0g}
\end{align}
where the $\nu$ are the bosonic Matsubara frequencies.
The matrix Green's function within BHF is \cite{Ozawa2012b}
\begin{align}
	&\mathbf{G}^{-1}(\mathbf{q},z) =
	\begin{pmatrix}
	z - A & -gn_0 & i\frac{\kappa}{m}q_y & 0
	\\
	-gn_0 & -z - A & 0 & i\frac{\kappa}{m}q_y
	\\
	-i\frac{\kappa}{m}q_y & 0 & z - B & 0
	\\
	0 & -i\frac{\kappa}{m}q_y & 0 & -z - D
	\end{pmatrix}, \label{greenmotion}
\end{align}
where
\begin{align}
	A(\mathbf{q}) & \equiv q^2/2m -\mu +g( 2n_0+2n_-+n_+) 
	\notag \\
	B(\mathbf{q}) & \equiv (2\boldsymbol\kappa + \mathbf{q})^2 / 2m -\mu +g( n_0+n_-+2n_+)
	\notag \\
	D(\mathbf{q}) & \equiv B(-\mathbf{q}). \label{abd}
\end{align}

Since the condensate transition is characterized by the infrared structure of the Green's functions,
in expanding in $n_0$ we consider only the $\nu = 0$ component in (\ref{dfdn0g}).
In evaluating Green's functions in (\ref{dfdn0g}), we approximate the $n_-$ and $n_+$  in $A$, $B$, and $D$ through (\ref{abd}) by $n(\mu)/2$, their value to  leading order in $n_0$; including the $n_0$ dependence of $n_\pm$ by solving the number equations self-consistently remains a task for the future. For $\kappa \neq 0$,
the right side of (\ref{dfdn0g}) can be expanded for small $gn_0 \ll   \kappa^2/m$ as 
\begin{align}
	\frac{\partial \mathcal{F}}{\partial n_0}
	=
	-\mu + \frac{3}{2}gn(\mu)
	+ X gn_0 + Y (gn_0)^2 + \cdots, \label{dfdn0p}
\end{align}
where
\begin{align}
	X(\mu, T) &\equiv 1 - \frac{4m^2gT}{\kappa}\alpha \left(\frac{\Delta \mu}{\epsilon_\kappa}\right),
	\notag \\
	Y(\mu, T) &\equiv \frac{4m^2 g T}{\kappa\epsilon_\kappa}\beta \left(\frac{\Delta \mu}{\epsilon_\kappa}\right), \label{defxy}
\end{align}
$\epsilon_\kappa \equiv \kappa^2/2m$, and $\alpha(x)$ and $\beta (x)$ are dimensionless functions.
Note that, if one replaces $g$ by $4\pi a/m$, the prefactor $4m^2g^2T$ becomes $32\pi^2a/\lambda^2$, where $\lambda$ is the thermal wavelength.

\begin{figure}[htbp]
\begin{center}
\includegraphics[width=8cm]{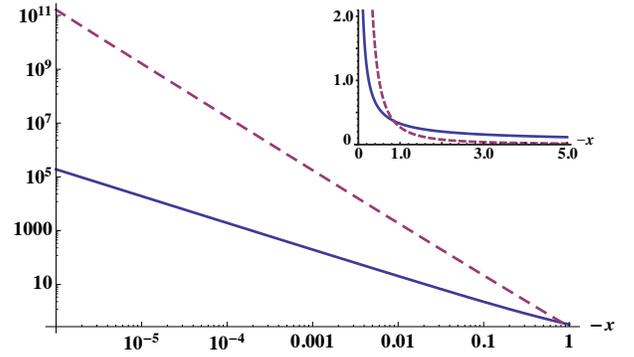}
\caption{Functions $\alpha (x)$ (solid line) and $\beta (x)$ (dashed line) [see Eq.~(\ref{defxy})] for $0>x>-1$.
The inset shows the behavior of the functions in the lower right corner, where $-x\, \simge\, 1$. 
 }
\label{alphabeta}
\end{center}
\end{figure}

Since $\Delta \mu < 0$, we need only consider $\alpha (x)$ and $\beta (x)$ with negative arguments;
by explicit calculation, one sees that $\alpha(x)$ and $\beta(x)$ are both positive there and monotonically increasing functions of $x$, with the asymptotic forms as $x \to 0^-$, 
\begin{align}
	\alpha (x) & \simeq -\frac{19}{32 \pi x}  \sim -\frac{0.19}{x},
	&
	\beta (x) & \sim \frac{0.16}{x^2},
	\label{asymp}
\end{align}
and approaching 0 as $x \to -\infty$.  The analytic form for $\alpha$ is derived from the $2G_{11}+G_{33}$ term in Eq.~(\ref{dfdn0g}); the contribution of the $G_{12}$ terms is small and numerically changes the coefficient from 0.19 to 0.20. Figure \ref{alphabeta} plots $\alpha (x)$ and $\beta (x)$, calculated numerically.

\begin{figure}[htbp]
\begin{center}
\includegraphics[width=8.5cm]{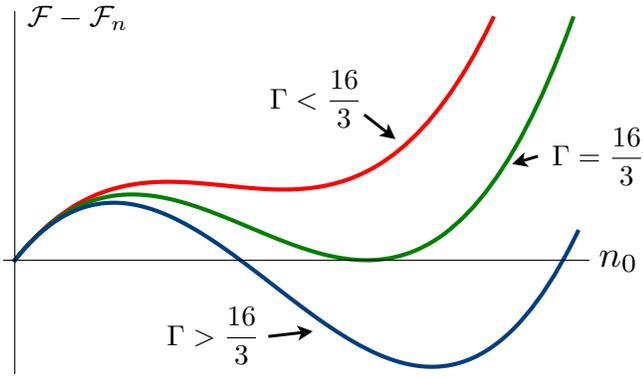}
\caption{The first order phase transition from the normal to condensed phase at $\Gamma= 16/3$. The lines show the Ginzburg-Landau free energy, measured with respect to the free energy of the normal phase.  The top line is for $T>T_c$, where
$\Gamma <  16/3$; the middle line is  at $T_c$, where $\Gamma =  16/3$; and the bottom line is for $T < T_c$, where 
$\Gamma >  16/3$.}
\label{gltransition}
\end{center}
\end{figure}

Integrating (\ref{dfdn0p}), we obtain the Ginzburg-Landau free energy up to third order in $n_0$:
\begin{align}
	\mathcal{F}(n_0)
	=
	\mathcal{F}_{\mathrm{n}}
	-\Delta \mu\, n_0
	+
	X\frac{g}{2}n_0^2
	+
	Y\frac{g^2}{3}n_0^3,
\end{align}
where $\mathcal{F}_{\mathrm{n}} = \mathcal{F}(n_0=0)$ is the free energy in the normal phase.
The coefficients of $n_0$ and $n_0^3$ are both positive, whereas, for a given $\mu > 0$, the coefficient $X$ of $n_0^2$ is negative at low $T$ and decreases continuously with decreasing temperature
 \footnote{To see that $X$ decreases with decreasing $T$, we use the asymptotic form (\ref{asymp}) for $\alpha$ together with Eq.~(\ref{neqapprox}) to derive the result
$$ \frac{d(\Delta\mu/T)}{dT}= -\frac{\mu/T^2}{\left[1- 3mg\kappa T/4\pi\Delta\mu\right]},   
$$
which is negative for positive $\mu$ as expected.}.
This change in the coefficient of $n_0^2$ drives a first order phase transition since, at sufficiently small $T$, 
the two conditions $\mathcal{F}(n_0) = \mathcal{F}_{\mathrm{n}}$ and $\partial{\mathcal F}(n_0)/\partial n_0 = 0$ become simultaneously satisfied;  this occurs when $X^2 = -16Y\Delta \mu / 3$.  At this temperature, the system undergoes a transition to the condensed phase.
Figure~\ref{gltransition} schematically shows how this transition takes place.
The combination $X^2/(-Y \Delta \mu) \equiv\Gamma$ monotonically decreases with $T$, as long as $X < 0$.
At the transition, $n_0$ jumps from zero to $|3X/4Yg| > 0$ on the condensate side.

{\em Transition temperature.}
Now, we estimate the transition temperature, assuming that the spin-orbit coupling strength is sufficiently large that $\epsilon_\kappa \gg |\Delta \mu|$. We will see that this condition is obeyed by typical experimental parameters.
Then, using the asymptotic forms (\ref{asymp}), the condition for the transition $\Gamma = 16/3$ becomes
\begin{align}
	-\frac{1}{2m\kappa g}\frac{\Delta \mu}{T_c} \frac{1}{0.16}
	\left[
	1 + 0.20\times 2m\kappa g\frac{T_c}{\Delta \mu}
	\right]^2
	\approx
	\frac{16}{3}.
	 \label{eqtm}
\end{align}
The left side depends only on the combination $\Delta \mu(T_c) / (2m\kappa g T_c)$; thus,
$\Delta \mu / (2m\kappa g T_c)$ takes a constant universal value.
Numerically solving (\ref{eqtm}), choosing the solution which makes $X < 0$, we find $\Delta \mu(T_c)/2m\kappa g T_c \sim -0.033$.
Then, using Eq.~(\ref{neqapprox}), we obtain the transition temperature, for  $2 m\kappa g \simle 1$, in terms of the normal state density, as
\begin{align}
	T_c \approx
	\frac{2\pi n(\mu)}{m\kappa}\frac{1}{|\ln (2m\kappa g)| + C}, \label{tcapprox}
\end{align}
where $C \sim 3.4$.
Thus, the transition temperature depends linearly on the density and as expected approaches zero as $g \to 0$ since there is no Bose condensate at nonzero temperature in the absence of interactions.

The jump in the condensate density at the transition is 
\begin{align}
	\frac{n_0}{n(\mu)}
	\sim
	\frac{0.32}{|\ln (2m\kappa g)|+C}, \label{n0approx}
\end{align} 
approaching zero as $g \to 0$ and increasing as $g$ increases.

As noted, expressions (\ref{tcapprox}) and (\ref{n0approx}) are valid when the spin-orbit coupling strength $\kappa$ is large compared with $\sqrt{mgn_0}$. When $\kappa = 0$, the system reduces to the ordinary two-component Bose gas without spin-orbit coupling and thus has a second order phase transition to a condensate phase at nonzero temperature.  Understanding the details of the quantum phase transition as $\kappa \to 0$ is beyond the purview of the current paper and is left for the future.

{\em Experimental feasibility.}
Finally, we mention accessing the transition experimentally. For orientation, we take typical current experimental values from Ref.~\cite{Lin2011}, which realized a gas of $^{87}$Rb with a mixture of Rashba and Dresselhaus spin-orbit coupling, with coupling strength $\kappa \sim \sqrt{2}\pi/800$nm and density of order $n \sim 10^{12}/\mathrm{cm}^3$.
Approximating the coupling $g$ by $4\pi a/m$, so that  $2m\kappa g \sim 0.69$, we obtain $T_c \sim 2~\mathrm{nK}$ with a jump in the condensate fraction at the transition $n_0 / n \sim 0.1$.
In addition, $T_c/\epsilon_\kappa \sim 0.019$ and $\Delta \mu/\epsilon_\kappa \sim 4.3 \times 10^{-4}$, putting the system in a regime where the approximations leading to (\ref{tcapprox}) are valid.
Future experiments could, depending on their specific configurations, be able to increase the transition temperature and the jump in the condensate density via increasing $\kappa$, $n$, or $a$.
Since a plane-wave condensate breaks rotational symmetry around the z-axis, the transition can be accompanied by formation of domains of condensates with different plane-wave momenta,
which is experimentally observable.
Including effects of trapping potentials and details of specific configurations of realizing artificial spin-orbit coupling is a work in progress.

\begin{acknowledgements}
This research was supported in part by NSF Grant No. PHY09-69790. G.B. is grateful to the Aspen Center for Physics, supported in part by NSF Grant No. PHY10-66293, where part of this work was carried out, and to Leon Balents and Matthew Fisher for helpful discussions. We also thank Jason Ho and Markus Holzmann for helpful comments and discussions.
\end{acknowledgements}


\begin{thebibliography}{99}

\bibitem{Lin2011} Y.-J. Lin, K. Jim\'{e}nez-Garc\'{\i}a, and I. B. Spielman, Nature (London) {\bf 471}, 83 (2011).

\bibitem{Williams2012} R. A. Williams, L. J. LeBlanc, K. Jim\'{e}nez-Garc\'{\i}a, M. C. Beeler, A. R. Perry, W. D. Phillips, and I. B. Spielman, Science {\bf 335}, 314 (2012).

\bibitem{Wang2012arXiv} P. Wang, Z.-Q. Yu, Z. Fu, J. Miao, L. Huang, S. Chai, H. Zhai, and J. Zhang, Phys. Rev. Lett. {\bf 109}, 095301 (2012).

\bibitem{Rashba1960} E. I. Rashba, Fiz. Tverd. Tela (Leningrad) \textbf{2}, 1224 (1960) [Sov. Phys. Solid State \textbf{2}, 1109 (1960)].

\bibitem{Dresselhaus1955} G. Dresselhaus, Phys. Rev. {\bf 100}, 580 (1955).

\bibitem{Ozawa2012b} T. Ozawa and G. Baym, Phys. Rev. Lett. {\bf 109}, 025301 (2012).

\bibitem{Barnett2012} R. Barnett, S. Powell, T. Gra\ss, M. Lewenstein, and S. Das Sarma, Phys. Rev. A {\bf 85}, 023615 (2012); Phys. Rev. A {\bf 85}, 049905(E) (2012).

\bibitem{Ho2011} T.-L. Ho and S. Zhang, Phys. Rev. Lett. {\bf 107}, 150403 (2011).

\bibitem{Wang2010} C. Wang, C. Gao, C.-M. Jian, and H. Zhai, Phys. Rev. Lett. {\bf 105}, 160403 (2010).

\bibitem{Zhai2012} H. Zhai, Int. J. Mod. Phys. B {\bf 26}, 1230001 (2012).

\bibitem{Ozawa2012a} T. Ozawa and G. Baym, Phys. Rev. A {\bf 85}, 013612 (2012).

\bibitem{Li2012} Y. Li, L. P. Pitaevskii, and S. Stringari, Phys. Rev. Lett. {\bf 108}, 225301 (2012).

\bibitem{Jian2011} C.-M. Jian and H. Zhai, Phys. Rev. B {\bf 84}, 060508(R) (2011).

\bibitem{Wu2011} C. Wu , I. Mondragon-Shem, and X.-F. Zhou, Chin. Phys. Lett. {\bf 28}, 097102 (2011).

\bibitem{Sinha2011} S. Sinha, R. Nath, and L. Santos, Phys. Rev. Lett. {\bf 107}, 270401 (2011).

\bibitem{Hu2012} H. Hu, B. Ramachandhran, H. Pu, and X.-J. Liu, Phys. Rev. Lett. {\bf 108}, 010402 (2012).

\bibitem{Xu2012} Z. F. Xu, Y. Kawaguchi, L. You, and M. Ueda, Phys. Rev. A {\bf 86}, 033628 (2012).

\bibitem{Ozawa2012c} T. Ozawa and G. Baym, Phys. Rev. A {\bf 85}, 063623 (2012).

\bibitem{Stanescu2008} T. D. Stanescu, B. Anderson, and V. Galitski, Phys. Rev. A {\bf 78}, 023616 (2008).

\bibitem{Holzmann2003} M. Holzmann and G. Baym, Phys. Rev. Lett. {\bf 90}, 040402 (2003); see also G. Baym and G. Grinstein, Phys. Rev. D {\bf 15}, 2897 (1977).

\bibitem{Brazovskii1975} S. A. Brazovskii, Zh. Eksp. Teor. Fiz. {\bf 68}, 175 (1975) [Sov. Phys. JETP {\bf 41}, 85 (1975)].

\bibitem{Jackeli2004} G. Jackeli and M. E. Zhitomirsky, Phys. Rev. Lett. {\bf 93}, 017201 (2004).

\bibitem{bkt-holzmann}  M. Holzmann, G. Baym, J.-P. Blaizot, and F. Lalo{\"e},
Proc. Natl. Acad. Sci. U.S.A. {\bf 104}, 1476 (2007).

\bibitem{Shenoy1997} V. B. Shenoy and T.-L. Ho, Phys. Rev. Lett. {\bf 80}, 3895 (1998).

\bibitem{Gopalakrishnan2011} S. Gopalakrishnan, A. Lamacraft, and P. M. Goldbart, Phys. Rev. A {\bf 84}, 061604(R) (2011).

\bibitem{Ozawa2011} T. Ozawa and G. Baym, Phys. Rev. A {\bf 84}, 043622 (2011).

\end{thebibliography}
\end{document}